\documentclass[twocolumn,aps,pra,groupedaddress,amsmath,amssymb,showpacs,floatfix]{revtex4}
\usepackage{graphicx}  
\usepackage{subfigure}
\begin{document}
\def\bar{\begin{eqnarray}}
\def\ear{\end{eqnarray}}
\def\beq{\begin{equation}}
\def\eeq{\end{equation}}
\newcommand{\degrees}{\ensuremath{^{\circ}}}
\title{Guided Modes of Elliptical Metamaterial Waveguides}
\author{Klaus Halterman}
\author{Simin Feng}
\author{P. L. Overfelt}
\affiliation{Physics and Computational Sciences, Research and Engineering Sciences Department, Naval Air Warfare Center,
China Lake, California 93555}
\date{\today}


\begin{abstract}
The propagation of guided electromagnetic waves in open elliptical metamaterial waveguide
structures is investigated. 
The waveguide contains a negative-index media core,
where the permittivity, $\varepsilon$ and permeability $\mu$
are negative over a given bandwidth.
The allowed mode spectrum for these structures
is numerically calculated by solving a dispersion relation
that is expressed in terms of Mathieu functions. By probing
certain regions of parameter space, we find
the possibility exists to have  
extremely localized waves that transmit along the surface of the waveguide.
\end{abstract}
\maketitle



Waveguides are  structures that are typically designed to transmit 
energy  along a specified trajectory with minimal attenuation and signal distortion.
When transmitting surface waves, this implies confining the traveling wave
within or adjacent to the waveguide walls \cite{kapany}.
Various avenues can be pursued when attempting to improve on a waveguides' capabilities, 
including modifying the constitutive effective material parameters of the guide.
The recent surge of interest in negative index of refraction materials \cite{pendry},
or negative index media (NIM), has prompted
a reanalysis of many conventional results for waveguide devices,
and a subsequent search for exotic transmission characteristics
when incorporating this particular composite media, or metamaterial, into
a variety of waveguide configurations. 

A crucial feature of NIM, is the frequency dispersion
of the permittivity, $\varepsilon$, and permeability, $\mu$,
with  $\varepsilon$ and $\mu$ simultaneously rendered negative over
a particular bandwidth. This results in
wave propagation in which the 
phase velocity and energy flow
of an electromagnetic wave
can be antiparallel, and also the possibility of
a negative index of refraction, proposed long ago \cite{ves}.
The
study of various  NIM based open waveguide structures and resonators 
during the past few years has 
demonstrated a number
of interesting effects:
it was shown \cite{eng} that a quasi 1-D bilayer resonator containing a NIM layer
can be substantially smaller than the usual cavity size due to phase cancellation.
For the case of a thin planar NIM waveguide, the TM mode can propagate 
for arbitrary widths and posses a single mode for slow waves \cite{eng}.
Guided TE modes were also shown to
have electric field profiles
containing nodes \cite{ilya},
and exhibit a sign-varying Poynting vector \cite{klaus,ilya}.
Similar results were reported in circular NIM fibers \cite{novitsky}.

Distortion of the circular dielectric waveguide into an elliptical guide,
while maintaining the cross sectional area, has been shown to reduce attenuation of 
the dominant mode \cite{yeh} and modal degeneracy, 
allowing for practical guiding of traveling electromagnetic waves.
Attenuation effects and power flow expressions were achieved 
for wave propagation in a surface wave transmission line with an elliptical cross section \cite{reng1}.
It was found that some modes in the guide have lower attenuation than the corresponding modes
in a circular guide.
The slow and fast hybrid mode spectrum in a metallic elliptical waveguide with a 
confocal dielectric lining
was also calculated \cite{reng2}, demonstrating the potential for a surface wave transmission device.
Moreover, the distribution of electromagnetic fields in a double layer elliptical waveguide
was calculated to first order and revealed in some cases, 
dispersion solutions where the power and phase velocity directions are antiparallel\cite{ray}.
Spurred by the recent theoretical and experimental advances in NIM waveguides and resonators,
we
examine in this paper,
open NIM waveguides with elliptical cross sections, affording a greater
flexibility in the parameter space
determined by the geometry and material constraints.
For a given eccentricity, $e$,
we calculate the permitted propagation constants over a
range of frequencies.
The allowed modes are separated into fast and slow propagating regions
of the dispersion diagram, where wave localization is
demonstrated in the form of electric field distributions.

\begin{figure}
\centering
\includegraphics[width=.45\textwidth]{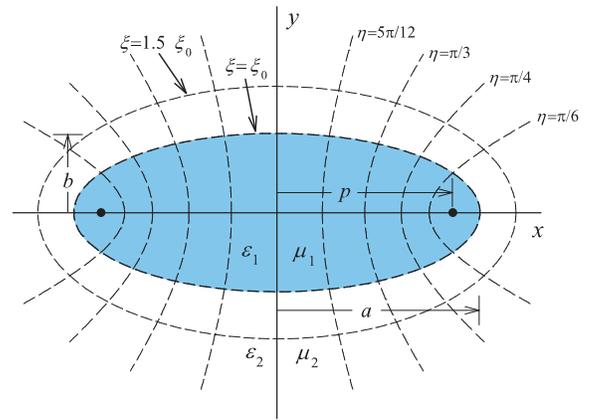}
\caption{Cross section of the open elliptical waveguide structure.
Within the  metamaterial core, $\varepsilon_1$ and $\mu_1$ are
in general both negative, and frequency dispersive. The outer region is air; $\varepsilon_2=1$ and $\mu_2=1$.}
\label{coord} 
\end{figure}
When searching for exact solutions to Maxwell's equations, it
is convenient to work in a coordinate system in which
the boundary of the structure coincides with
one of the coordinates being held constant.
It is thus appropriate for the geometry under consideration to
work
in an orthogonal elliptical coordinate system 
described by the coordinates $\xi$ and $\eta$, 
depicted in Fig.~\ref{coord}. Elliptical coordinates are related to their 
rectangular counterparts via
$x=p\cosh\xi\cos\eta$ and 
$y=p\sinh\xi\sin\eta$, 
for  $0\leq\xi<\infty$ and $0\leq\eta\leq2\pi$,
where $p\equiv(a^2-b^2)^{1/2}$ is the semifocal length of the ellipse expressed
in terms of the
semimajor and semiminor axes $a$ and $b$ respectively. 
The waveguide boundary is
located at $\xi=\xi_{0}$, and hence  $a=p \cosh \xi_{0}$ and $b=p\sinh\xi_{0}$.
It then follows that the eccentricity, $e$, is written $e=[1-(b/a)^2]^{1/2}$=$1/\cosh\xi_0$,
such that $0\leq e<1$, with $e=0$ corresponding to a circular cross section.
In order to obtain the electric
($\bf E$) and magnetic ($\bf H$) fields, we must solve
the vector Helmholtz wave equation obtained from Maxwell's equations. 
This yields three coupled
second order differential equations for either $\bf E$ or $\bf H$.
The waveguide along the axial ($z$) direction is translationally invariant
and thus the form of the equation governing
the longitudinal $E_z$ (or $H_z$) is identical to the scalar
Helmholtz equation in elliptic coordinates, permitting a 
semi-analytic and more tractable
solution. 
The fields are assumed to vary harmonically in time and propagation occurs in the
positive $z$ direction.
The wave equation then reduces to the following form:
\begin{align}
\label{helm}
\frac{1}{h^2}\left(\frac{\partial^2}{\partial\xi^2}+\frac{\partial^2}{\partial\eta^2}\right)
E_{z i}(\xi,\eta) +k_i^2 E_{zi}(\xi,\eta)
=0,
\end{align}
where the index $i$ denotes the region,
$ \sqrt{2}h\equiv p\sqrt{\cosh(2 \xi)-\cos(2\eta)}$, 
$k_i^2=\mu_i(\omega) \varepsilon_i(\omega) \omega^2/c^2-k_z^2$,
and for the unattenuated modes of interest here, 
$k_z$ is the real propagation constant along the $z$-direction.
We have also suppressed  the usual $\exp(i k_z z-i\omega t)$ factor.  
A similar equation exists for $H_z$.
Due to the cylindrical symmetry, the transverse field components 
$(E_{\eta i},E_{\xi i})$ and $(H_{\eta i},H_{\xi i})$ can be
determined \cite{jacko} from $E_{z i}$ and $H_{z i}$:
\begin{subequations}
\begin{align}
E_{\eta i}(\xi,\eta)&=\frac{i k_z}{k_i^2 h}
\left(\frac{\partial E_{z i}}{\partial \eta}-\mu_i\frac{k_0}{k_z}\frac{\partial H_{z i}}{\partial \xi}\right), \\
E_{\xi i}(\xi,\eta)&=\frac{i k_z}{k_i^2h}
\left(\frac{\partial E_{z i}}{\partial \xi}+\mu_i\frac{k_0}{k_z}\frac{\partial H_{z i}}{\partial \eta}\right), \\
H_{\eta i}(\xi,\eta)&=\frac{i k_z}{k_i^2h}
\left(\frac{\partial H_{z i}}{\partial \eta}+\epsilon_i\frac{k_0}{k_z}\frac{\partial E_{z i}}{\partial \xi}\right), \\
H_{\xi i}(\xi,\eta)&=\frac{i k_z}{k_i^2h}
\left(\frac{\partial H_{z i}}{\partial \xi}-\epsilon_i\frac{k_0}{k_z}\frac{\partial E_{z i}}{\partial \eta}\right).
\end{align}\label{heta}
\end{subequations}
We proceed by inserting ${E_{z i}(\xi,\eta)}=U_i(\xi)V_i(\eta)$ into Eq.~(\ref{helm}), splitting it into 
two ordinary differential equations
with separation constant $\Lambda$:
\begin{subequations}
\begin{align}
\frac{d^2 V_i(\eta)}{d\eta^2} + (\Lambda-2q_i\cos2\eta)V_i(\eta) &=0,\label{mat} \\
\frac{d^2 U_i(\xi)}{d\xi^2} -(\Lambda- 2q_i\cosh2\xi)U_i(\xi) &=0,\label{modmat}
\end{align}
\label{matall}
\end{subequations}
where
$q_1= (k_1 p/2)^2$, and $q_2= -(k_2 p/2)^2$.

The angular  Mathieu equation [Eq.~(\ref{mat})] 
describes the angular variation of the field around the ellipse.
Two linearly independent periodic solutions exist, 
akin to the trigonometric $\sin$ and $\cos$
functions: the even and odd  angular Mathieu functions, denoted
$ ce_{n}(\eta;q_i)$ and $se_{n}(\eta;q_i)$, respectively.
For the elliptical waveguide problem, the angular Mathieu functions
must be periodic with period $2\pi$ and of integer order,
otherwise the solution set is nonperiodic and can be unstable \cite{mex}.
We  can thus expand the angular Mathieu functions in a Fourier series, where
the expansion coefficients 
can be found 
recursively
by substituting the expansions 
back into Eq.~(\ref{mat}).
The parameter $\Lambda$ found in Eqs.~(\ref{matall}) can be solved
using a method of continued fractions {\cite{mex}}. The orthogonality and normalization
relations of the angular Mathieu functions are given by Eq.~(\ref{ortho}).
The angular Mathieu functions are also periodic in $\eta$ only for special
characteristic values of $\Lambda$, 
denoted here as $a_n(q_i)$  for the even solutions and $b_n(q_i)$ for the odd ones.

The radial Mathieu equation (\ref{modmat}),
admits two general types of  solutions:
the radial Mathieu functions
of the first and second kind.
The solutions of the first kind are divided into
even and odd functions denoted as, $Ce_n(\xi;q_i)$ and $Se_n(\xi;q_i)$ respectively.
The calculation of
the characteristics numbers and the Fourier coefficients
are exactly the same as for the angular functions.  In
identifying a convenient series expansion to compute
the radial Mathieu functions, a number of techniques are applicable\cite{abram}.
We found, in agreement with past works, that the Bessel $J_n$ product 
series are the most stable
basis of functions to use \cite{mcg}. 
Not surprisingly, the $Ce_n$ and $Se_n$ functions coalesce into $J_n$, 
as the elliptical cross section degenerates to a circular one.
The 
Mathieu functions of the second kind for $q_i>0$ are denoted $Fey_n(\xi;q_i)$ (even), and  $Gey_n(\xi;q_i)$ (odd),
and
are calculated 
similarly, except the series expansion involve products of both the Bessel $Y_n$ and $J_n$ functions.
The functions $Fey_n$ and $Gey_n$ are analogous 
to the
Bessel functions of the second kind, $Y_n$, in circular coordinates.
When $q_i<0$, $Fey_n$ and $Gey_n$
transform into $Fek_{n}(\xi;q_i)$ and $Gek_{n}(\xi;q_i)$ (analogous to the Bessel $K$ functions), 
and they are related \cite{mcg}, e.g., for even order, 
$2 Fek_{2n}(\xi;-q_i)={(-1)^{n}}\left[ -Fey_{2n}(\xi+i \pi/2;q_i) +i Ce_{2n}(\xi+i\pi/2;q_i)\right]$.



For a general 
cylindrical open guide with circular cross section,
pure TE or TM modes exist only for symmetrical electromagnetic  fields, i.e.,
independent of the azimuthal angle.
For an
open elliptical waveguide however, 
the reduction in symmetry forces
the electromagnetic modes in a given region to be hybrid in that
both the longitudinal fields, $E_{z i}$ and $H_{z i}$ exist simultaneously ($i$ denotes the region). 
Based on the discussion above, it is clear that
field solutions in an elliptical domain are split into
even and odd components: $E_{z i}\rightarrow \lbrace E^{\rm e}_{z i}, E^{\rm o}_{z i}\rbrace$
and $H_{z i}\rightarrow \lbrace H^{\rm e}_{z i}, H^{\rm o}_{z i}\rbrace$.
The procedure adopted here consists of writing
$H^{\rm e}_{z i}$ ($H^{\rm o}_{z i}$)  waves 
as products involving even (odd)  Mathieu functions, and
$E^{\rm e}_{z i}$ ($E^{\rm o}_{z i}$) waves in terms of odd (even) Mathieu functions \cite{yeh}.
The complete solution is then
expanded
as products of angular and radial Mathieu functions of the requisite parity, and that
obey the radiative condition. 
The boundary conditions constrain 
the types of Mathieu functions
used in the expansions \cite{mcg}, as
we require ``stationary waves" in the transverse $\eta$ and $\xi$ direction.
Within each region, we 
thus seek
solutions of the form,
\begin{subequations} \label{fields1}
\begin{align}
E_{z i}^{{\rm e}}(\xi,\eta,z)&=\sum_{m=1}^{\infty} a_{m i} A_{m i}(\xi;q_i)  se_m(\eta;q_i)e^{i k_z z}, \label{ez1}\\
H_{z i}^{\rm e}(\xi,\eta,z)&=\sum_{m=0}^{\infty} b_{m i} B_{m i}(\xi;q_i) ce_m(\eta;q_i)e^{i k_z z}, \label{ez2}
\end{align}
\end{subequations} 
where $a_{m i}$ and $b_{m i}$ 
are constants, 
and the coefficients
$A_{m i}(\xi;q_i)$, and $B_{m i}(\xi;q_i)$ 
contain the ``radial" dependence to the fields, given below.
Without loss of generality,
the methodology used here
will focus on waves of 
even parity, as 
the procedure for odd waves is similar.

Within the waveguide region (see Fig.~\ref{coord}), the solutions are the 
radial Mathieu functions of the first kind,
$A_{m 1}(\xi;q_1)=Se_m(\xi;q_1)$, and $B_{m 1}(\xi;q_1)=Ce_m(\xi;q_1)$.
For the guided modes of interest here, the fields 
must decay from the surface at $\xi=\xi_0$, and therefore
within the surrounding medium we have,
$A_{m 2}(\xi;q_2)\equiv Gek_m(\xi;-q_2)$ and $B_{m 2}(\xi;q_2)\equiv Fek_m(\xi;-q_2)$.
With this requirement on the fields, and
for a given set of material and geometrical parameters,
a restricted number of waveguide modes exist.
To determine the allowed modes, 
the tangential ${\bf E}$ and ${\bf H}$ 
fields
are
matched at the boundary $\xi=\xi_0$ separating
the two media. 
The $\eta$-dependence is integrated out 
by making use of the orthogonality properties of the angular Mathieu
functions (see Appendix). We then cast the boundary matched equations
into a linear equation system containing in principle, an infinite hierarchy of Mathieu functions.
The higher order Mathieu functions 
arise from
the lack of
one-to-one correspondence between 
the angular Mathieu functions in regions with differing material parameters: 
the arguments of the angular Mathieu functions depends on the parameter $q_i$, which
in turn 
depends on $\varepsilon$ and $\mu$ of the relative media. 
This is in contrast to a circular waveguide, where the angular dependence is a function
of only integer multiples of the azimuthal coordinate, $m \phi$.

In order to find the nontrivial solution, 
the problem of finding the allowed modes thus amounts to finding where 
the associated determinant of the linear equation system vanishes
over a range of frequencies, propagation constants and ellipticities.
The methodology we shall discuss  is valid for hybrid ${\rm HE}_{11}$ 
or ${\rm EH}_{11}$ (the first letter
represents the dominant field) modes \cite{yeh}, with small $e$, and frequencies
corresponding to small $|\varepsilon_1-\varepsilon_2|$ (and small $|\mu_1-\mu_2|$).
Under these conditions, the expansions in Eqs.~(\ref{fields1}) can be limited
to the first few terms. Taking for example, the two lowest order terms in the outer region, yields the
following even mode dispersion relation,
\begin{align}
\label{tran}
&\Biggl(\frac{\varepsilon_1}{q_1}
\frac{{Se^\prime_1}(q_1)}{Se_1(q_1)}
+\frac{\varepsilon_2}{q_2}
\frac{{Gek^\prime_1}(-q_2)}{Gek_1(-q_2)}\Biggr)
\Biggl(\frac{\mu_1}{q_1}
\frac{{Ce^\prime_1}(q_1)}{Ce_1(q_1)}
+\frac{\mu_2}{q_2}
\frac{{Fek^\prime_1}(-q_2)}{Fek_1(-q_2)}\Biggr) \nonumber \\
&+\frac{k^2_z}{k_0^2}\frac{1}{\alpha_{11}\beta_{11}}\Bigl[\frac{1}{q_1 q_2}\left(\psi_{11} C_1 + \tau_{11} C_2\right)
+\frac{1}{q_2^2} C_1 C_2  \nonumber \\
&+\frac{1}{q_1^2}\tau_{11}\psi_{11}\Bigr] 
=0,
\end{align}
where ${k}_0 = \omega/c$, $C_1\equiv -\beta_{11}\gamma_{11}-\gamma_{13}\beta_{31}$, and
$C_2\equiv \alpha_{11} \gamma_{11} + \alpha_{31}\gamma_{31}$.
The quantities $\alpha_{m n}, \beta_{m n}, \tau_{m n}, \psi_{m n}$, and $\gamma_{m n}$, outlined in the Appendix,
arise from the integrals of products of overlapping angular Mathieu functions.
Writing the dispersion relation explicitly in this way reduces computational time considerably by
avoiding unstable numerical determinants
of large matrices and the associated multiple function calls
to higher order
Mathieu functions.
The odd mode spectrum is easily obtained
via the interchange
$\varepsilon_i \leftrightarrow -\mu_i$.
As the elliptical cross section degenerates into 
a circular one, $C_1\rightarrow 1$,
$C_2\rightarrow -1$, 
and the Mathieu functions appropriately transform
into their corresponding cylindrical Bessel functions: 
$\lbrace Ce^\prime_1 (q_1)/Ce_1(q_1), Se^\prime_1 (q_1)/Se_1(q_1)\rbrace \rightarrow u {J^\prime_1}(u)/ J_1(u)$, and 
$\lbrace Fek^\prime_1 (-q_2)/Fek_1(-q_2), Gek^\prime_1 (-q_2)/Gek_1(-q_2)\rbrace \rightarrow v {K^\prime_1}(v)/K_1(v)$. 
In this limit, the dispersion relation (\ref{tran})
reduces to the familiar characteristic equation
for a circular waveguide:
\begin{align}
\label{circle_disp}
\Biggl(\varepsilon_1
\frac{{J_1}^\prime(u)}{u J_1(u)}
+\varepsilon_2
\frac{{K_1}^\prime(v)}{v K_1(v)}\Biggr)
\Biggl(\mu_1
&\frac{{J_1}^\prime(u)}{u J_1(u)}
+\mu_2
\frac{{K_1}^\prime(v)}{v K_1(v)}\Biggr) \nonumber \\
&-\left[\frac{k_z}{k_0} \frac{(u^2+v^2)}{u^2 v^2}\right]^2=0,
\end{align}
where 
$u^2= k_1^2 a^2$ and $v^2= -k_2^2 a^2$.

\label{results}

\begin{figure}
\centering
\includegraphics[width=.5\textwidth]{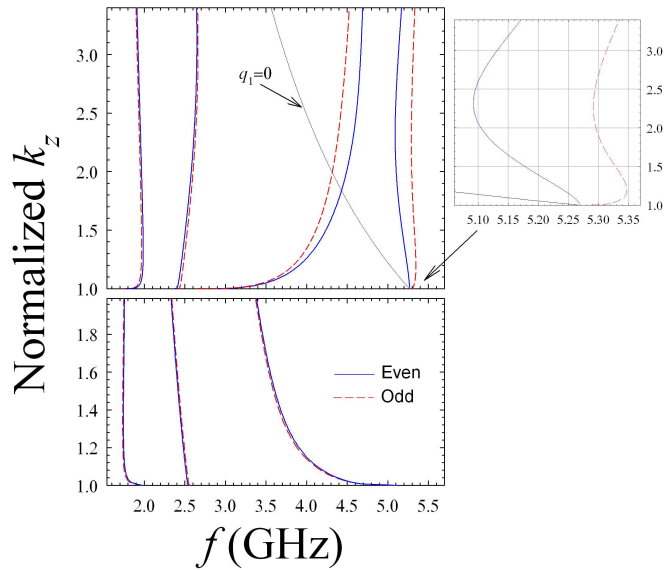} \\
\caption{Surface wave spectrum for the even and odd hybrid modes of an 
elliptical 
waveguide filled with NIM and embedded in air (top panel). 
The bottom panel is the same configuration with positive
The propagation constant, $k_z$, is normalized
by the free space wavenumber $k_0$ and is plotted as a function of
the operating frequency for a cross section corresponding
to an ellipticity of $e=0.44$. 
We take the characteristic frequencies
in the material dispersion to be $\omega_p=8 (2\pi)$ GHz, and $\omega_m=7 (2\pi)$ GHz,
which corresponds to $\mu_1,\epsilon_1\leq-1$ 
within the frequency range, $f\leq \omega_m/(2\pi \sqrt{2})$. 
The mode spectra illustrates the slow and fast modes, separated by the 
$q_1=0$ curve. If the NIM core is replaced with media corresponding to
positive $\varepsilon_1$ and $\mu_1$ (bottom panel), the slow wave solution is absent and
the curves vary in an opposite sense.
}
\label{fig0} 
\end{figure}

The surface wave dispersion relation (\ref{tran}) is
a function of the parameters, $k_z$, $\omega$, and 
eccentricity, $e$; only particular combinations of these quantities that satisfy Eq.~(\ref{tran})
are allowed mode solutions. 
The  hybrid waves that are explored here can possess even and odd components, and
are denoted appropriately in subsequent figures. We present the propagation 
constant in terms of the convenient dimensionless ratio, $k_z/k_0$,
and the frequency units are all in GHz.
The waveguide cross section is assumed to not deviate greatly
from that of a circular guide, reflected in moderate values of $e$. 
The
permittivity and permeability in the NIM regions, $\varepsilon_1$ and $\mu_1$, respectively, have the
frequency dispersive 
form,
$\varepsilon_1=1-\left({\omega_p}/{\omega}\right)^2$ and $\mu_1=1-\left({\omega_m}/{\omega}\right)^2$,
where $\omega_p$ and $\omega_m$ are the effective electrical and magnetic plasma frequencies \cite{klaus}, respectively.
Thus for $\omega\leq \omega_p/\sqrt{2}$, we have $\varepsilon_1\leq-1$.
The study here is concerned with frequency regions of parameter space where 
$\mu_1$ and $\varepsilon_1$ are simultaneously negative.

Determining the allowed modes typically involves
holding
the semimajor and semiminor axes of the waveguide (and hence $e$) fixed, and then
scanning Eq.~(\ref{tran}) over $k_z$ and $\omega$.
Any sign change that occurs
signifies a zero crossing that can be pinpointed 
through an iterative root finding process.
Other variable combinations may be used, depending on the 
parameter study.
In the top panel of Fig.~\ref{fig0} we show
the mode spectra found by solving Eq.~(\ref{tran}), and its odd counterpart, for a
waveguide with eccentricity, $e=0.44$.
For clarity, only
four neighboring sets of dispersion curves within the given frequency window are shown.
Two sets reside completely beneath the $q_1=0$ curve, one set entirely
outside of it, and another set that traverses both regions.
For this geometry and range of frequencies,
the paired even-odd mode solutions follow similar trends,
separating at higher frequencies.
The solutions to the Helmholtz equation (Eq.~\ref{helm}) depend on the
eccentricity, $e$, of the elliptical guide through the parameters $q_1$ and $q_2$.
We found as the ellipse flattens,
more splitting occurs between the even and odd modes.
In general, each branch of the even and odd modal curves coalesce
at cutoff, where the propagation constant
approaches the free space value ($k_z\rightarrow k_0$). 
The dispersion diagram portrays the allowed electromagnetic wave solutions
that travel along the guide and it elucidates important information
regarding the possible localization characteristics of guided modes.
In particular, the first three sets of dispersion curves that satisfy
$\sqrt{\varepsilon_2\mu_2} \leq {k}_z/k_0\leq \sqrt{\varepsilon_1\mu_1}$
correspond to conventional surface waves.
Within this parameter space region, the phase velocity of guided waves, $v_p$, exceeds the phase velocity of waves in a homogeneous
bulk medium, i.e., $v_p>c/\sqrt{\varepsilon_1\mu_1}$. 
These {\it fast} wave solutions
to Maxwell's equations will decay in the air region,
but not necessarily in the waveguide core. Indeed, as
the structure increases in size, or
as $f$ decreases (increasing $|\epsilon_1|$), the 
electric and magnetic fields inside the waveguide
oscillate with shorter wavelengths.
At cutoff, $k_2\rightarrow 0$ (thus
$q_2 \rightarrow 0$), and the ensuing decay length increases outside the
guiding surface. This causes a significant
portion of the energy flow to occur in the air region, where 
the fields can then become less sensitive to the relevant geometrical parameters, such as the ratio $a/b$.

We see in Fig. 2, that for the case of a NIM core, there are also solutions that reside outside of the $q_1=0$
boundary. These {\it slow} wave solutions are noticeably absent when $\varepsilon_1$ and $\mu_1$ are strictly 
positive (see bottom panel). 
The presence of NIM sets up a non-oscillatory field profile that 
rapidly decays outside the guide, allowing for the possibility of 
guided modes that are more localized to the surface,
akin to surface plasmon waves on metal surfaces studied long ago.
Note that
as $k_z$ increases, $q_2$ typically increases, in which case
the radial Mathieu functions decay sharply,
confining the field more to the waveguide.
The slow wave solutions seen in the inset of Fig.~\ref{fig0} have dispersion characteristics 
that depend strongly on 
frequency. Near cutoff, these modes demonstrate
that the direction of the group velocity, $d\omega/d k_z$,
evaluated at a particular $k_z$, can differ among the even and odd wave solutions.
This is related to the (time-averaged) local energy-density flow  along $z$, 
$S_{z i}=(1/2)\Re(E_{\xi i} H^*_{\eta i} - E_{\eta i} H^*_{\xi i})$.
An excited wave with
the appropriate frequency,
can reverse direction between layers,
a hallmark of the peculiar wave propagation that can arise in NIM guiding structures. 
To address the relationship between net energy flow in the system and group velocity,
\begin{figure}
\includegraphics[width=.4\textwidth]{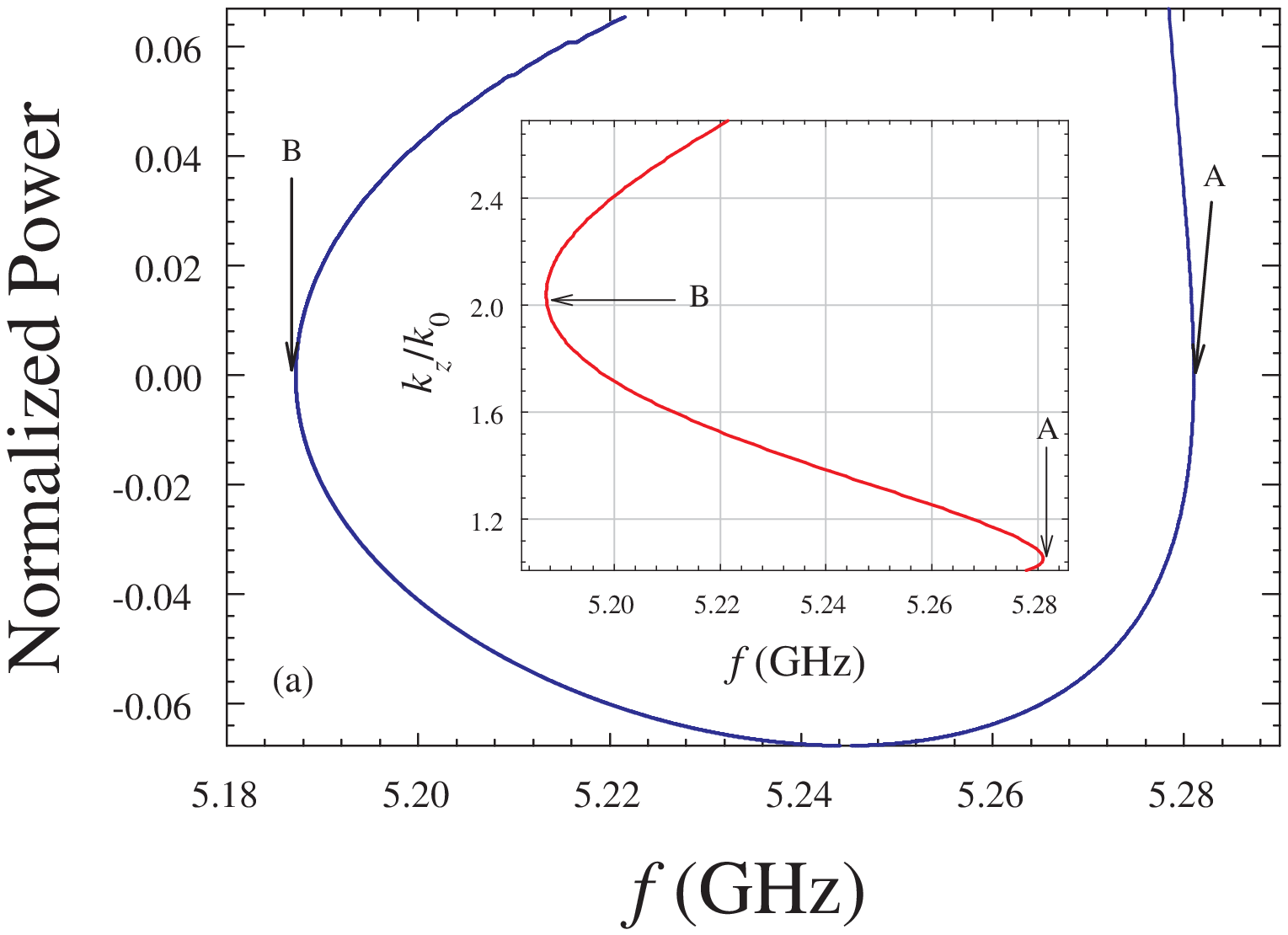}\\
\includegraphics[width=.4\textwidth]{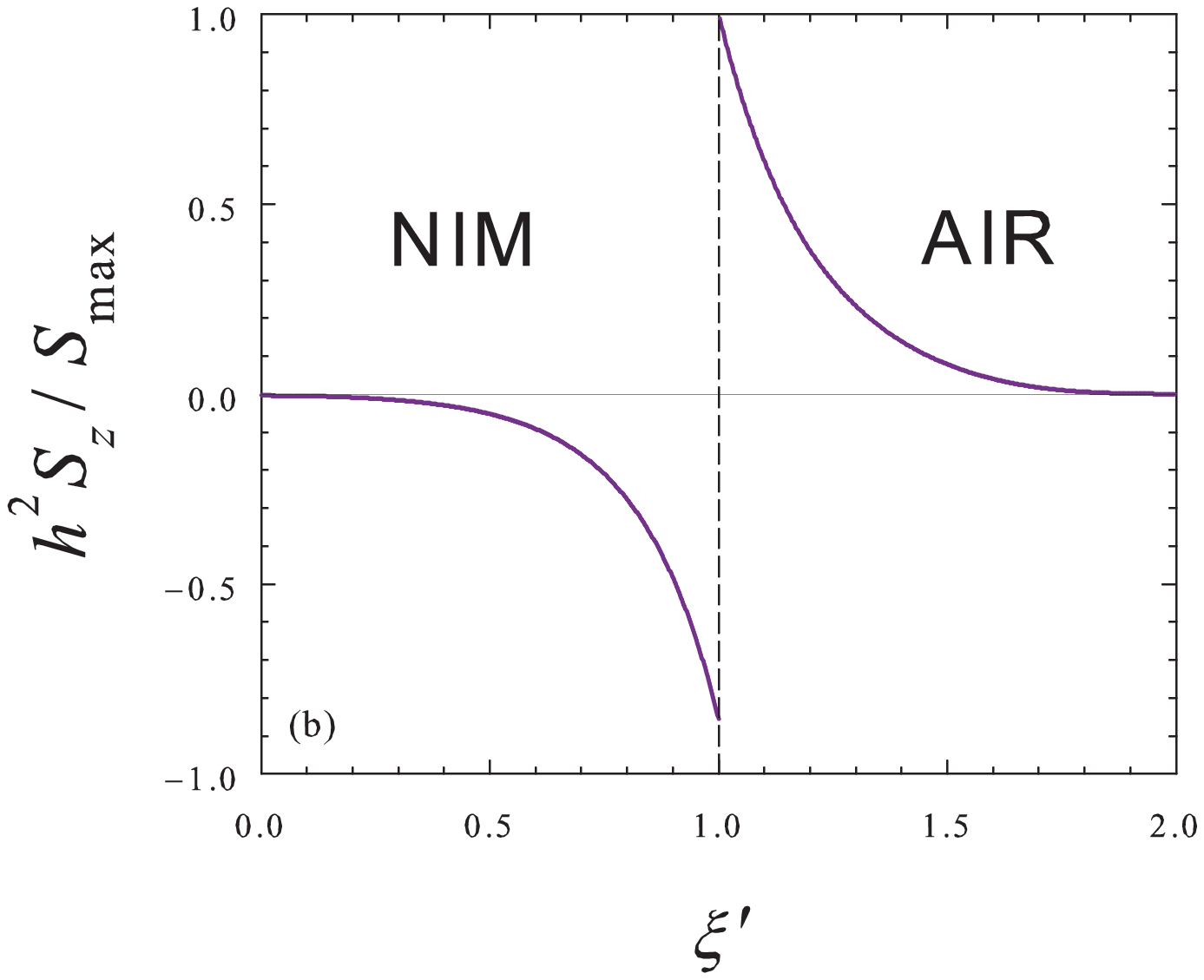}
\caption{(a) The total power, $P=P_1+P_2$ (normalized by $|P_1|+|P_2|$) versus the frequency
for (odd) slow wave modes. The eccentricity, $e$, is set at $e=0.59$.
The propagation constant, $k_z$, increases in going from $\rm A \rightarrow B$,
in accordance with the mode dispersion diagram in the inset.
In (b) energy-density reversal is shown by means of the Poynting vector, $S_z$,
normalized by its maximum, and as a function of the dimensionless coordinate, $\xi'\equiv\xi/\xi_0$. 
The dashed vertical line identifies the NIM-air boundary.
The frequency is fixed at $f=5.28$ GHz, and  $k_z=1.1$.
The spatial range is for the angle
$\eta=\pi/6$. }
\label{power} 
\end{figure}
we show in Fig.~\ref{power}(a), the total power, $P$ as a function
of the calculated mode frequency, found by summing
the power flow in the waveguide ($P_1$) and air ($P_2$) regions.
The power through a given cross sectional area, $A$, was calculated by integrating the 
$z$ component of the Poynting vector over $A$,
\begin{equation}
P_i = \iint_A S_{z i} h^2 d\eta d\xi,\qquad i=1,2,
\end{equation}
where $h$ is the usual coordinate scale factor.
The frequencies used in determining $P$ 
are governed by the dispersion curve, shown in the inset of Fig.~\ref{power}(a).
The arrows label points where 
$d\omega/d k_z$ is zero 
in the dispersion diagram, and correlate with
zero net power flow in the system, i.e., 
$P_1=-P_2$. 
In general, we see from the figure that the direction
of net power flow coincides with the sign of $d\omega/dk_z$.
A sign change in this slope 
causes the dispersion curves to bend back  in the $k_z-\omega$ plane, related to
the $P_1$ and $P_2$ sign difference that can yield a sum ranging from positive to
negative, depending on their relative values. 
The bottom panel (b) shows the spatial dependence 
of the energy-density flow and its associated reversal in going from NIM to air.

\begin{figure}
\includegraphics[width=.4\textwidth]{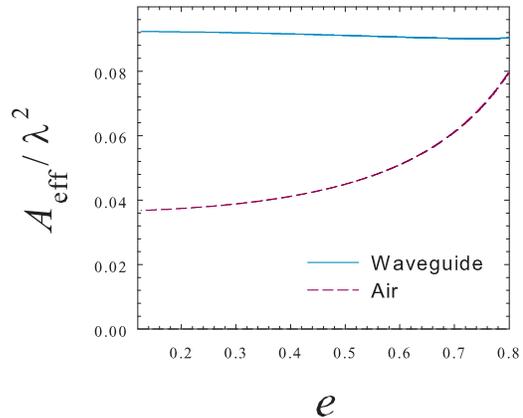}\\
\caption{The slow wave effective mode area, $A_{\rm eff}$ 
(normalized by the wavelength, $\lambda^2$) as a function
of the eccentricity,
$e$. 
}
\label{modevol} 
\end{figure}

We now characterize mode localization 
with
an effective mode area, $A_{\rm eff}$, defined as the ratio of the electromagnetic 
energy to the maximum value of the energy density,
\begin{equation}
A_{\rm eff} \equiv \frac{\iint_A U(\xi,\eta) h^2 d\eta d\xi}{U_{\rm max}(\xi,\eta)},
\end{equation}
where  the energy density, $U$, 
for frequency dispersive materials is defined
as \cite{shitz},
\begin{equation}
U(\xi,\eta)=\frac{1}{8\pi}\left[\frac{d(\omega \varepsilon)}{d\omega} |{\bf E}|^2 +\frac{d(\omega\mu)}{d\omega} |{\bf H}|^2\right].
\end{equation}
The normalized effective mode area for slow waves near cutoff ($k_z/k_0=1.01$) is shown in Fig.~\ref{modevol} as a function of the eccentricity, $e$.
The mode frequencies calculated from Eq.~(\ref{tran}) are a weak function
of $e$ over the range shown.
We see that $A_{\rm eff}$ in the air region tends to decrease
as the guide becomes more circular, while within the guide, the effective area is nearly constant. 
Thus, as a circular guide is slightly distorted into an ellipse,
the exhibited mode localization properties within the NIM structure remain rather robust.

\begin{figure}
\includegraphics[width=.22\textwidth]{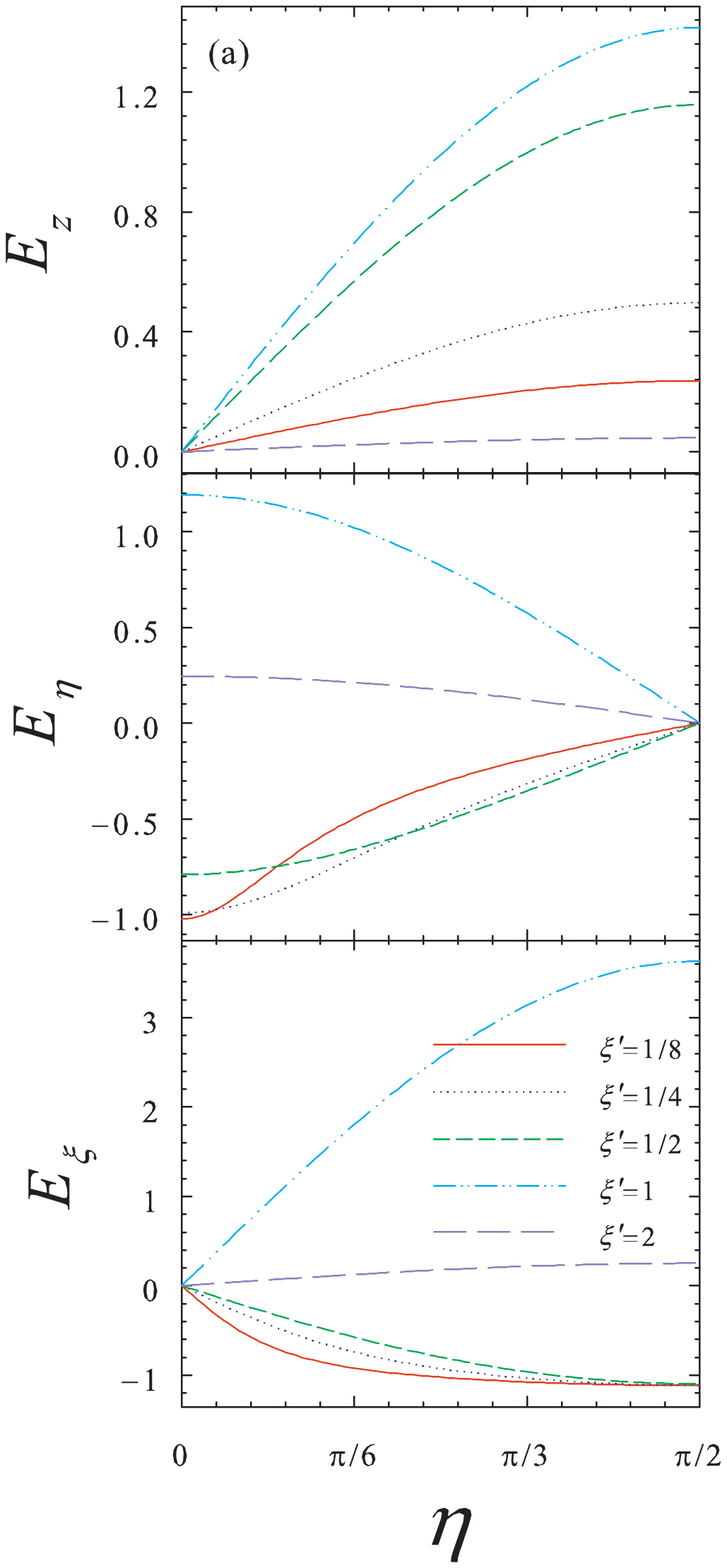}
\includegraphics[width=.22\textwidth]{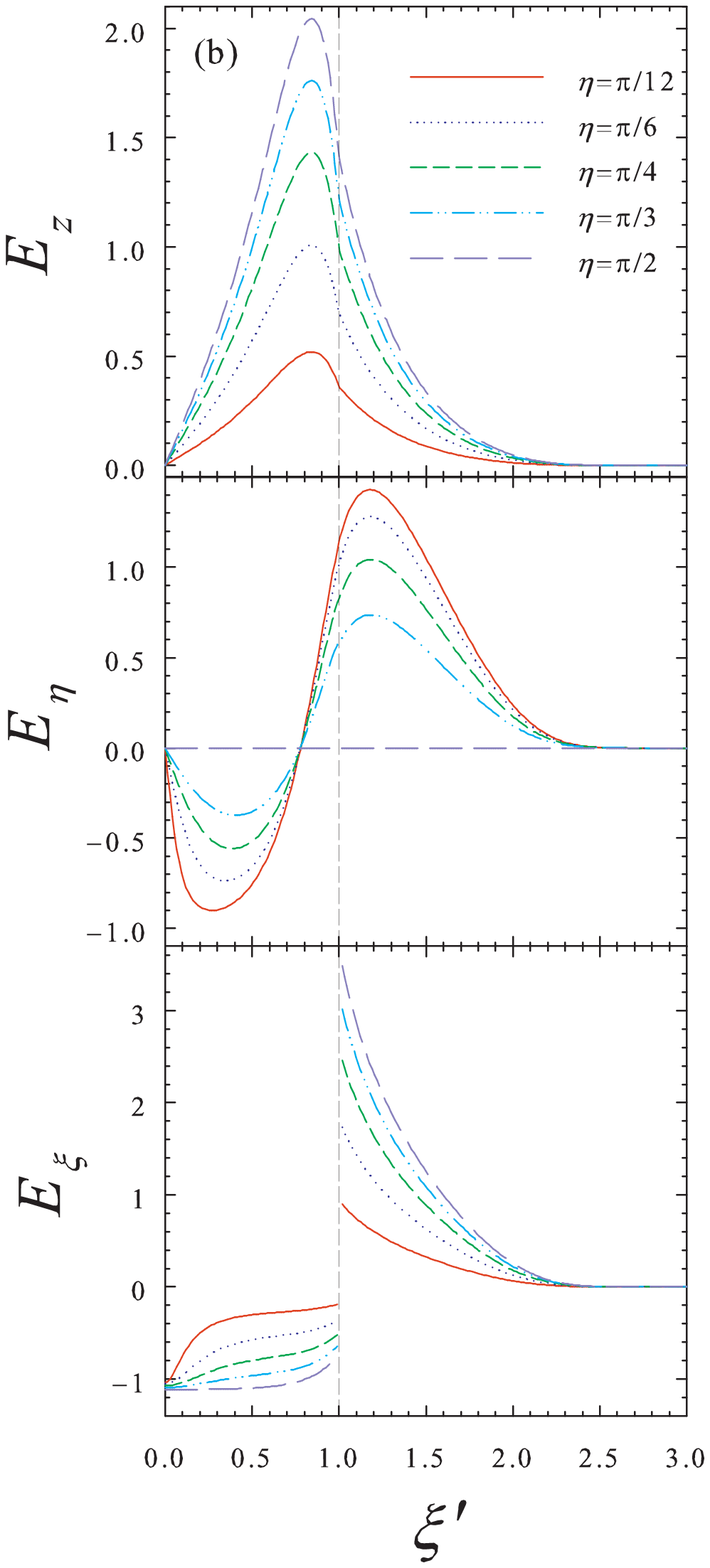}
\caption{The normalized electric field distributions for even wave modes near cutoff in an elliptical
waveguide as a function of (a) angular coordinate, $\eta$, and (b) normalized radial coordinate, $\xi'\equiv\xi/\xi_0$. 
The frequency is set at $3.25$ GHz, and  the ellipticity, $e$, has the value $e=0.14$,
corresponding to $\xi_0=2.65$. The vertical dashed line at $\xi'=1$ identifies
the waveguide boundary.
}
\label{field1} 
\end{figure}

To gain insight into the transmission properties of
an elliptical NIM waveguide excited by a particular source,
the spatial and angular features of the EM field components is essential.
We therefore show in Fig.~\ref{field1}, the three components of the $\bf E$ field near cutoff,
and at a frequency corresponding to $\varepsilon_1=-5.04$ and $\mu_1=-3.63$. The cross sectional area, $\pi a b$,
of the waveguide is held fixed with $e=0.14$.
The left panel, Fig.~\ref{field1}(a), illustrates the normalized ${\bf E}$ field
as a function of $\eta$, for $5$ different values of the normalized coordinate, $\xi'\equiv\xi/\xi_0$. 
The right set of figures, Fig.~\ref{field1}(b),
exhibits the normalized electric field as a function of $\xi'$ for
5 different $\eta$.
Parenthetically, in comparing field distributions with a circular waveguide,
one can scale the coordinates (e.g. for $\eta=\pi/2$): $\xi \rightarrow p\sinh\xi$. 
From panel (a), we see that the longitudinal $E_z$ and
``radial" $E_\xi$ components have the expected behavior
along the semimajor axis, since the $\eta$ dependence to those fields involve $se_m(\eta;q_i)$ 
and $ce^\prime_m(\eta;q_i)$ terms, which vanish at $\eta=0$. Likewise, $E_\eta$ is comprised of products
involving $ce_m(\eta;q_i)$ and $se^\prime_m(\eta;q_i)$ functions, and
hence tends toward zero for positions along the semiminor axis ($\eta=\pi/2$).
Turning to the $\xi$ dependence in panel (b), it is evident
that within the waveguide region, for $\xi=0$ (along the line $x'=p\cos \eta$),
the $z$ and $\eta$ components to the field vanish, while, $E_\xi$ has
its maximum there. It is apparent that on average, each of the components 
are similar in magnitude, with $E_\xi$ dominating slightly over the others in
some instances. Another distinguishing feature among
the components is the shifting of the peak intensity of the field patterns:
$E_z$ reaches its peak value inside of the
waveguide, $E_\eta$ has its largest value just outside the core boundary, while $E_\xi$ peaks out
along the waveguide walls ($\xi'=1$), after which it
undergoes a discontinuous transition. This behavior is
consistent with the boundary condition $\Delta E_\xi(\xi_0) = (1-\varepsilon_1/\varepsilon_2) E_{\xi 1}(\xi_0)$.
\begin{figure}
\includegraphics[width=.22\textwidth]{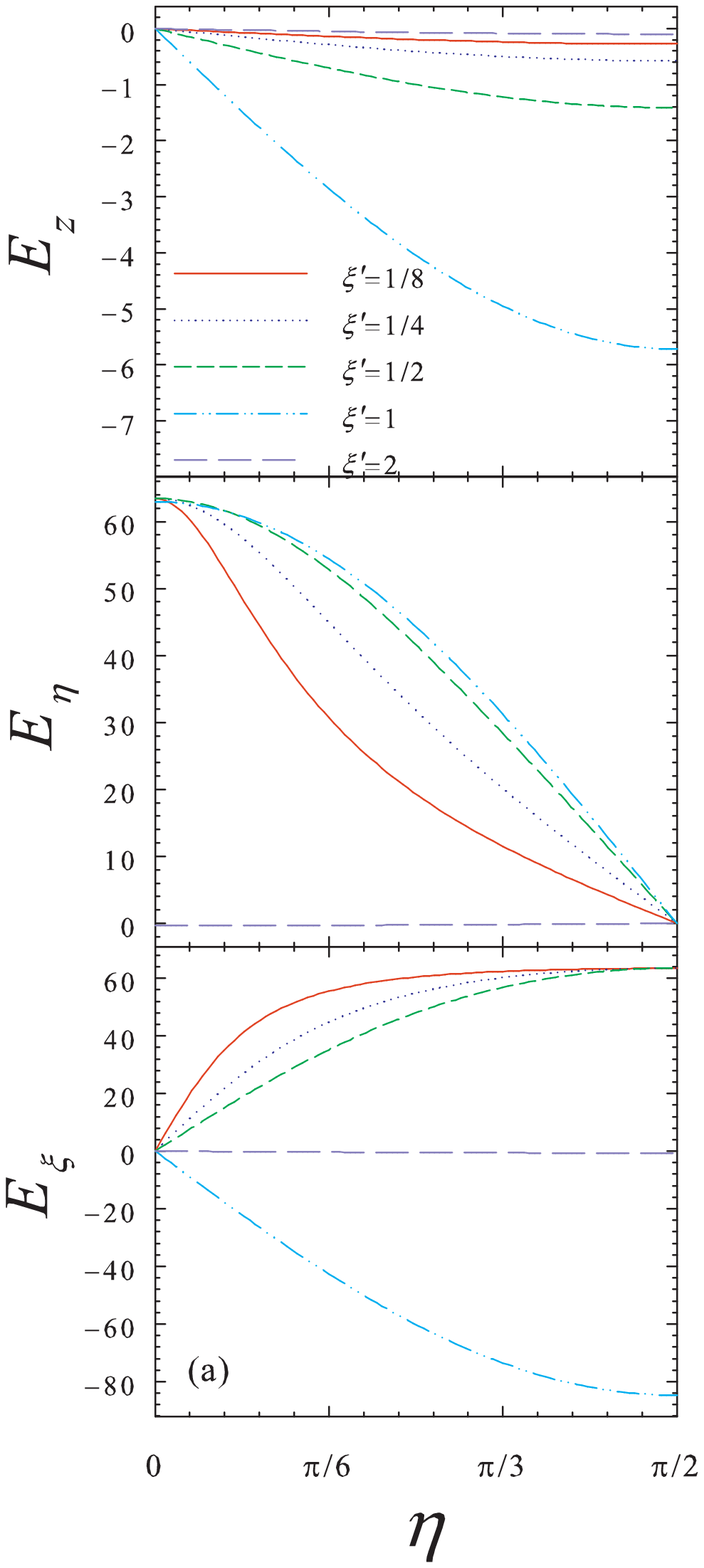}
\includegraphics[width=.22\textwidth]{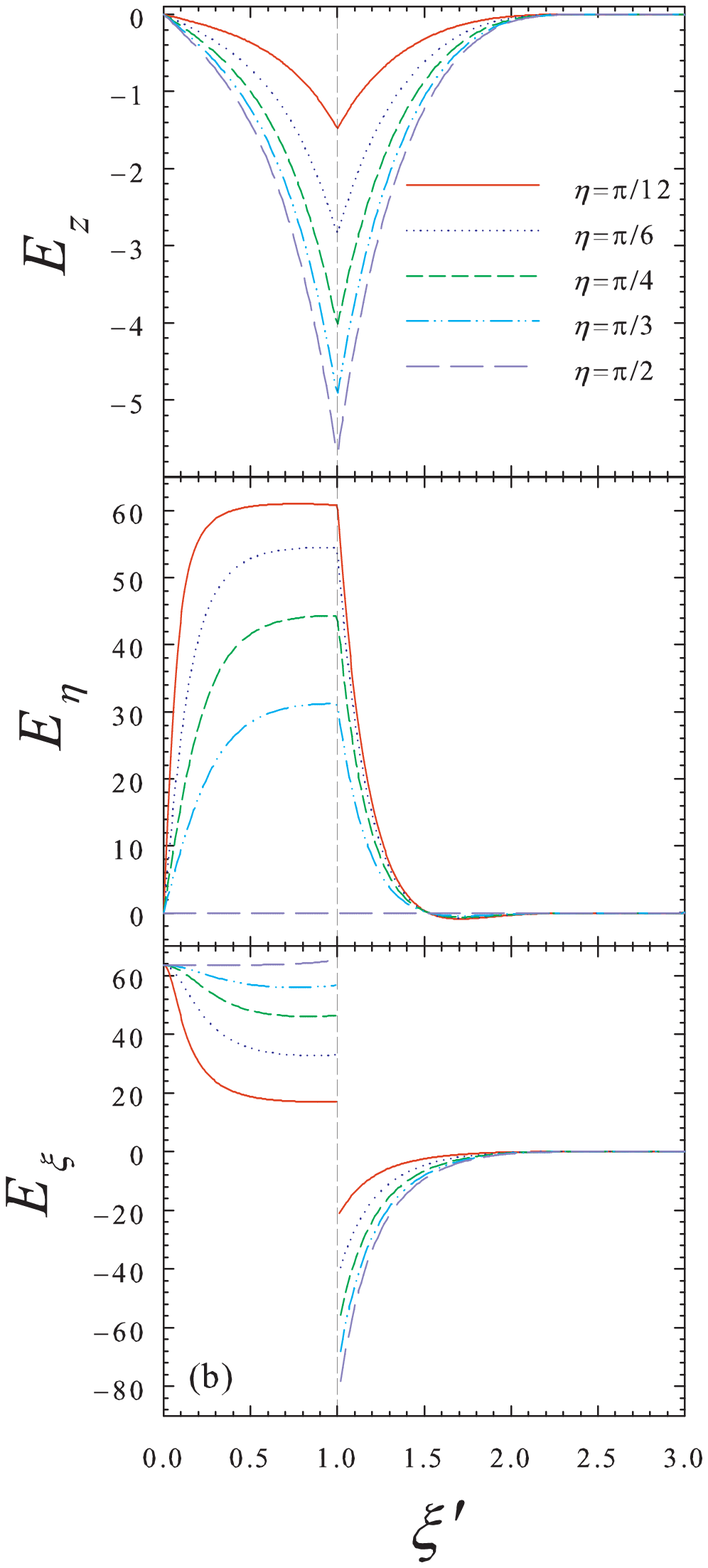}
\caption{The spatial and angular variation of
the electric field, with the same parameters as in Fig.~\ref{field1}, except at
the higher frequency, $f=5.28$ GHz. These localized slow waves
decay away from the outer side of the interface rapidly, consistent with
dispersion curves of Fig.~\ref{fig0}.
}
\label{field2} 
\end{figure}

To explore the possibility of field localization
in the slow wave regime, Fig.~\ref{field2}
illustrates the localized slow wave solutions as a function of $\eta$ and $\xi$.
The frequency chosen, $5.28$ GHz, lies just outside the $q_1=0$ curve at $k_z/k_0\approx 1.01$,
and approximate frequency (in GHz),
\begin{equation}
f\approx 
\frac{\omega_p \omega_m}{2\pi\sqrt{\omega_m^2+\omega_p^2}}\left(1-\frac{[(k_z/k_0)^2-1] \omega_m^2\omega_p^2}{
2(\omega_m^2+\omega_p^2)^2}\right)=5.26.
\end{equation}
Examining  the interior of the waveguide, Fig.~\ref{field2}(b)
is consistent with Fig.~\ref{field1}(b)
at $\xi=0$, where only  
$E_\xi$ survives before declining towards the interface.
As $\eta\rightarrow\pi/2$, $E_\xi$ becomes
more weakly dependent upon the coordinate $\xi$, while
if $\eta\rightarrow 0$, $E_\xi$ and $E_z$ vanish and $E_\eta$ approaches
a constant value within the guide. 
The ${\bf E}$ field components transverse to the direction
of energy flow, $E_\eta$ and $E_\xi$,  clearly dominate here,
and thus the behavior of these particular modes
is quite relevant
in the determination of waveguide transmission
capabilities. 
It is further evident from Fig.~\ref{field2} (b) that the length scale
of field decay in the air region
is at times shorter, demonstrating 
that the possibility exists to tailor the guide or feed line
in a way that transmits ultra-localized waves.
\begin{figure}
\includegraphics[width=.4\textwidth]{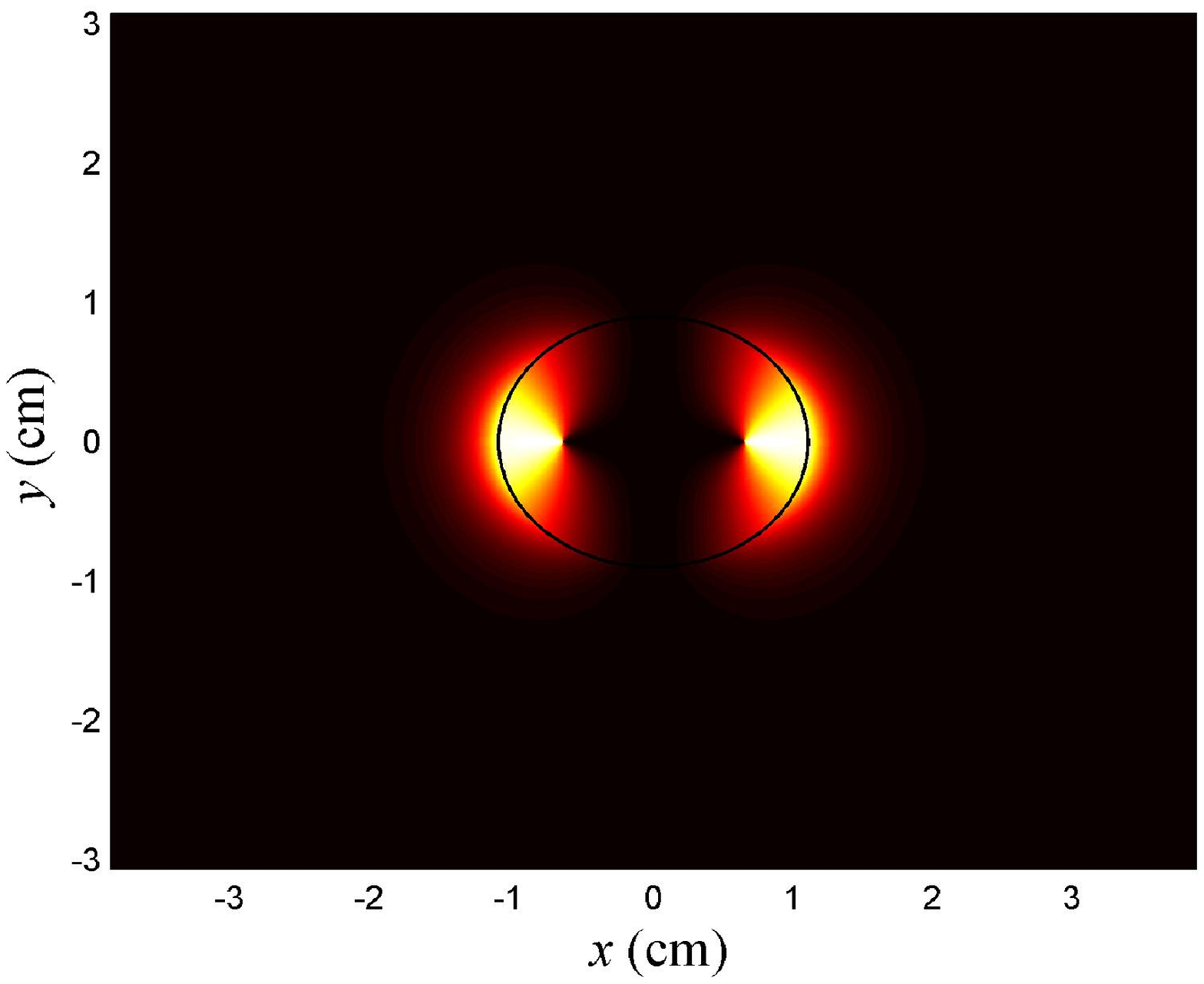}\\
\includegraphics[width=.4\textwidth]{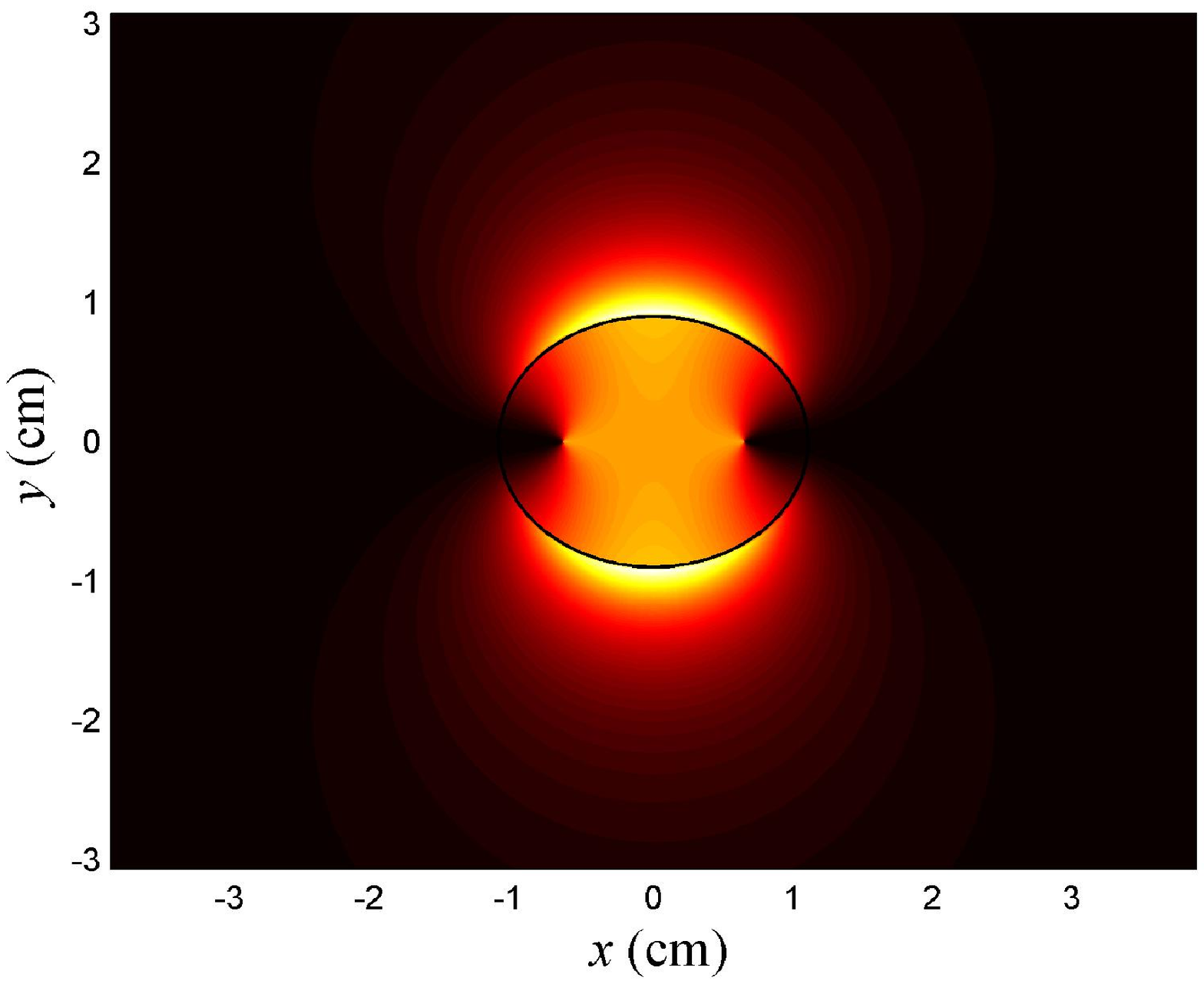}
\caption{Contour plots of the even transverse electric field amplitudes for $e=0.59$ and $f=5.3$ GHz.
The top and bottom panels correspond to $E_\eta$ and  $E_\xi$ respectively.
Bright
areas indicate larger field intensities. 
Each component is normalized to its respective maximum for clarity.
}
\label{field_contour} 
\end{figure}
The dominant transverse electric field profiles 
mapped onto a Cartesian coordinate system is shown in Fig.~\ref{field_contour}.
The 2-D  contour plots are consistent with the field patterns exhibited in
Fig.~\ref{field2}.

\vspace{-.01in}
In conclusion we have shown that elliptical waveguides with NIM
can support both fast and slow wave modes.
The power flow in the system was shown to have direct correlations
with the group velocity: points on the dispersion curves where the group velocity is zero
corresponded to zero net power flow through the entire structure.
The Poynting vector was shown to reverse when crossing the boundary between air and NIM.
The dispersion relation was shown to admit localized solutions that
retain their characteristics under moderate variations of the eccentricity.

\appendix
%

\section{Overlap Integrals of Angular Mathieu Functions}
\label{rel}
When matching the tangential fields at the boundary,
the  orthogonality properties of the angular Mathieu functions 
give rise to
several
overlap integrals, given as,
\begin{subequations}
\label{ortho}
\begin{align}
\alpha_{m n}&=\frac{1}{\pi}\int_0^{2\pi} d\eta\,ce_m(\eta;-q_2) ce_n(\eta;q_1), \\
\beta_{m n}&=\frac{1}{\pi}\int_0^{2\pi} d\eta\,se_m(\eta;-q_2) se_n(\eta;q_1),\\
\tau_{m n}&=\frac{1}{\pi}\int_0^{2\pi} d\eta\, se^{\prime}_m(\eta;q_1) ce_n(\eta;-q_2), \\
\psi_{m n}&=\frac{1}{\pi}\int_0^{2\pi} d\eta\,ce^{\prime}_m(\eta;q_1) se_n(\eta;-q_2), \\
\gamma_{m n}&=\frac{1}{\pi}\int_0^{2\pi} d\eta\,ce^{\prime}_m(\eta;-q_2) se_n(\eta;-q_2).
\end{align}
\end{subequations}
When $q_1=q_2$, $\beta_{m n}=\delta_{m n}$, and $\alpha_{m n}=\delta_{m n}$.
In the limiting case of the ellipse reducing to a circle,
then we also have $\tau_{m n}\rightarrow m \delta_{m n}$
and $\psi_{m n}\rightarrow -m\delta_{m n}$, and $\gamma_{m n} \rightarrow -m\delta_{m n}$.
In all cases, the integrals are zero if the $m$ is even {\it and} 
$n$ is odd or vice versa, due to the symmetry properties of the products of
periodic Mathieu functions.

\vspace{-.26in}
\acknowledgments
\vspace{-.19in}
This project is
funded in part by the Office of Naval Research
(ONR) In-House Laboratory Independent Research (ILIR) Program and by a grant of HPC resources from 
the Arctic Region Supercomputing Center at the University of Alaska Fairbanks 
as part of the Department of Defense High Performance Computing Modernization 
Program. 


\begin{thebibliography}{99} 
\bibitem{kapany} N. S. Kapany and J. J. Burke, {\it Optical Waveguides}. (Academic Press, 1972).
\bibitem{pendry}  J. B. Pendry,  Phys. Rev. Lett. {\bf85}, 3966 (2000).
\bibitem{ves}  V. G. Veselago,  Sov. Phys. Usp. {\bf10}, 509 (1968).
\bibitem{eng} N. Engheta, IEEEE Ant. and Wire. Prop. Lett. {\bf 1}, 10 (2002);
A. Alu and N. Engheta, IEEE Trans. on Microw. Th. and Tech. {\bf 52}, 199 (2004).
\bibitem{ilya}  I.V. Shadrivov, A.A. Sukhorukov, and Y.S. Kivshar, Phys. Rev. E. {\bf 67}, 057602 (2003).
\bibitem{klaus}  K. Halterman, J. M. Elson, and P. L. Overfelt,  Opt. Express {\bf11}, 521 (2003).
\bibitem{novitsky} A. V. Novitsky and L. M. Barkovsky, J. Opt. A: Pure
Appl. Opt {\bf 7}, S51-S56 (2005).
\bibitem{ray} S.B. Rayevskiy, L.G. Simkina, and V. YA. Smorgonskiy,  Radiotekh elektron+, 985 (1972).
\bibitem{yeh} C. Yeh, J. Appl. Phys. {\bf 33}, 3235 (1962).
\bibitem{jacko} J.D. Jackson, {\it Classical Electrodynamics, 2nd ed.},
        pg. 341, (Wiley, New York, 1975).
\bibitem{abram}{\it Handbook of Mathematical Functions}, 
edited by  M. Abramowitz  and A. Stegun (U.S. GPO, Washington, D.C. 1964).
\bibitem{mcg} McLachlan, N. W.,
{\it Theory and Applications of Mathieu Functions}, New York: Dover, 1964.
\bibitem{reng1} S. R. Rengarajan and J. E. Lewis, IEEE T. Microw. Theory {\bf 28}, 1085 (1980).  
\bibitem{reng2} S. R. Rengarajan and J. E. Lewis, IEEE T. Microw. Theory {\bf 28}, 1089 (1980).  
\bibitem{mex} J. C. Guti\'{e}rrez-Vega, R. M. Rodr\'{i}guez-Dagnino, M. A. Meneses-Nava,and S. Ch\'{a}vez-Cerda,
Am. J. Phys. {\bf 71}, 233 (2003).
\bibitem{shitz}       L.D.Landau, E.M.Lifshitz   \hskip 2mm
                        {\it Theoretical Physics : Electrodynamics of Continuous Media}, 
			Oxford, London, New York : Pergamon Press, 1960. 
\end{thebibliography}
\end{document}